\documentclass[12pt]{article}

\usepackage{graphicx, epsfig}
\usepackage{color}
\usepackage{enumerate}
\newcommand{\beq}{\begin{equation}}
\newcommand{\eeq}{\end{equation}}
\newcommand{\bea}{\begin{eqnarray}}
\newcommand{\eea}{\end{eqnarray}}

\newcommand{\half}{\frac{1}{2}}

\newcommand{\Mbh}{M_{\rm BH}}
\newcommand{\Mpl}{M_{\rm Pl}}

\newcommand{\Mpbh}{M_{\rm PBH}}
\newcommand{\done}{(1+1)}
\newcommand{\dtwo}{(2+1)}
\newcommand{\dthree}{(3+1)}

\begin{document}

\begin{center}	
\vskip 2cm
{\bf \LARGE Primordial Black Hole Evaporation and Spontaneous Dimensional Reduction}
\vskip 1cm
{\large J. R. Mureika}\\
Department of Physics\\
Loyola Marymount University\\
Los Angeles, CA~~90045\\
Email: jmureika@lmu.edu

\end{center}

\vskip 1cm

{\noindent{\bf Abstract} \\
\noindent Several different approaches to quantum gravity suggest the effective dimension of spacetime reduces from four to two near the Planck scale.  In light of such evidence, this letter re-examines the thermodynamics of primordial black holes (PBHs) in specific lower-dimensional gravitational models.  Unlike in four dimensions, $\done$-D black holes radiate with power $P \sim \Mbh^2$, while it is known no $(2+1)-$D (BTZ) black holes can exist in a non-anti-deSitter universe.  This has important relevance to the PBH population size and distribution, and consequently on cosmological evolution scenarios.  The number of PBHs that have evaporated to present day is estimated, assuming they account for all dark matter.  Entropy conservation during dimensional transition imposes additional constraints.  If the cosmological constant is non-negative, no black holes can exist in the $(2+1)$-dimensional epoch, and consequently a $(1+1)$-dimensional black hole will evolve to become a new type of remnant.  Although these results are conjectural and likely model-dependent, they open new questions about the viability of PBHs as dark matter candidates. }\\

\section{Introduction}
It has long been known theories of gravitation have a much simpler formulation in $\dtwo$-D \cite{gott1,gott2,reznik,btz1,btz2,reznik2,hussain,clement,carlipbook,ortiz} and $\done$-D \cite{collas,rajeev,jackiw,brown,mann1,mann2,mannsik,manndan,mannross,mann3,mannliouville,kummer2,mann4,balbinot1,balbinot2,kummer,lineland,grumiller2}, where associated quantum theories are exactly solvable \cite{rajeev}.   A resurgence of interest in lower-dimensional physics has been spurred by a confluence of evidence that the effective dimensionality of spacetime may depend on the energy scale at which interactions take place \cite{vd1,jrmds1,jrmds2,vd2,vd3,Shirkov:2010sh,Carlip:2009km,nieves,obukhov,modesto,calcagni4,garcia,Hexia,rinaldi1,rinaldi2,ref2}.  Instead of revealing extra dimensions at very short distances \cite{add,rs}, it is conceivable that the number of spatial dimensions decreases as the Planck length is approached.    

Dynamical or spontaneous dimension reduction has been studied in various contexts, mostly focusing on the energy-dependence of the spacetime's spectral dimension $d_s$.  The latter is the effective dimension seen by a diffusion process on the manifold over some time $\sigma$, characterized by a probability return function $P(\sigma)$ \cite{cdt,horava}.  The spectral dimension is formally defined as $d_s = -2\frac{d\log P(\sigma)}{d \log\sigma}$, which for flat space is $d_s=4$ \cite{cdt}, indicating this quantity is a  probe of the underlying geometry.  The causal dynamical triangulation approach was the first to demonstrate the spectral dimension decreases to $d_s = 2$ as the energy scale increases  \cite{cdt}.  This effect is replicated in a noncommutative-inspired geometry \cite{lmpn}, as well as through the anisotropic scaling factors in Lifshitz gravity \cite{horava}.   In each model, the described mechanism ``hides'' structure of the manifold at scales approaching the quantum regime, turning gravity into an effective lower-dimensional theory. 

From a string theory perspective, it has been shown that an energy-dependent dimension emerges from a smooth transformation of a three-brane to a one-brane \cite{maldecena}.  Similarly, this idea has been extended to model a three-brane as a collection of one-branes at every point \cite{myers}.   Alternate dimensional reduction scenarios include fractal spacetimes \cite{calcagni4,piero0,calcagni1,calcagni2,calcagni3,calcagni5}, with additional approaches concerning new techniques in gauge coupling unification \cite{Shirkov:2010sh} and a strong coupling expansion of the Wheeler-DeWitt equation \cite{Carlip:2009km}. 

A geometric dimensional reduction framework was recently proposed wherein a $(d+1)$-dimensional spacetime is a recursive lattice-network of lower-dimensional substructures \cite{jrmds1,jrmds2,vd1,vd2,vd3}.  Each has a fundamental length scale $L_k$ that becomes relevant at the energy $E_k \sim L_k^{-1}$.  This concept naturally addresses the hierarchy problem, and provides a range of phenomenological signatures --  including dimensionally-dependent scattering cross-sections and gravitational wave frequency thresholds -- that could be observable in present or future experiments.  The idea is motivationally-similar to, but formally distinct from, the cascading DGP scenarios previously discussed in the literature \cite{dgp1,dgp2}.

A lower-dimensional Planckian arena for gravity is thus natural and attractive.  It is therefore important to fully understand the roles of the spectral and geometric dimensions as they relate to gravitational phenomenology.  Since the characteristics of spacetime are unknown at quantum scales, one is tempted to take advantage of this ambiguity and interpret the spectral dimension as the geometric dimension of the manifold.  This presents several intriguing questions: is the universe itself effectively lower-dimensional at high energies?  If so, how does the transition from one dimension to another affect the dominant physics, and ideally is it possible to observe evidence of such transitions?

Whatever the underlying framework, it will be assumed that the quantum geometry is described semiclassically by an effective $\done$-D or $\dtwo$-D metric.   In this letter, the former case will be represented by a dimensionally-reduced limit of Einstein gravity, and the latter will employ the three-dimensional BTZ metric.  Primordial black holes play a critical role in a range of early-universe processes, from baryogenesis \cite{baryo1,baryo2,baryo3} to large-scale structure formation \cite{lss1,lss2,lss3,afshordi}, and even potentially determining the entropy content of the universe \cite{uentropy}.  Since PBHs are possible dark matter candidates \cite{afshordi,carrhawk,carr2,carr3,frampton0,frampton,pierorobb}, understanding their evolution and abundances in dimensionally-reduced spacetimes can shed new light on this dilemma of modern cosmology.  

\section{Lower dimensional black hole thermodynamics}
A number of models have addressed gravity in two-dimensional spacetimes, all of which require the additional presence of a coupled scalar field (see {\it e.g.} \cite{kummer} for a comprehensive review).    It has been demonstrated that such general models will exhibit slightly different temperatures depending in part on the nature of the dilaton coupling.  A generic dilaton gravitational theory in two dimensions  can be derived from the action
\beq
S = \frac{1}{2} \int d^2x\sqrt{-g} e^{-2\phi} \left[R + 4a(\nabla \phi)^2 + Be^{2(1-a-b)\phi}\right]
\label{gen2daction}
\eeq
where the coefficients $a,b,$ and $B$ depend on the model in question (see \cite{kummer} for details).  For minimally-coupled fields, one finds the Hawking temperature to be
\beq
T(\alpha) \sim \Mbh^\alpha~~~,\alpha = \frac{a-1}{a}~~.
\label{minscalar1}
\eeq
so a variety of possible temperature profiles are possible depending on the value of $\alpha$.  

For the purposes of the present discussion, however, the theory of choice is one whose action is \cite{mannross}
\beq
S_2 = \int d^2x~\sqrt{-g} ( \psi R-\frac{1}{2}(\nabla \psi)^2 +{\cal L}_m-2\Lambda)~~,
\eeq
which can be derived as a dimensionally-reduced form of $D$-dimensional Einstein gravity.  A strength of this model -- and hence the rationale for its use in this study -- is that it is the best classical and semiclassical approximation for general relativity in the 2-D limit \cite{mannross,mann3,mann4}\footnote{Reference \cite{grumiller2} also notes the favorability of Liouville gravity, whose solutions and metric structure are virtually identical.}.  On variation, the dilaton decouples from the background and one obtains 
\beq
R - \Lambda = 8\pi G_{1} T~~;~~~\nabla_b T^{ab} = 0
\label{11efe}
\eeq
as the effective field equations.  This model guarantees a conserved stress-energy tensor, which is a desired consequence that enhances the traditional Jackiw action \cite{jackiw} (corresponding to ($a=0,b=1$) in (\ref{gen2daction}), with the general transformation $e^{-2\phi}\rightarrow \psi$).  This theory also has a one-dimensional Newtonian limit, and can be generalized to the case of a $(1+1)$-dimensional non-commutative geometry \cite{jrmpn}.

The solution to (\ref{11efe}) is 
\beq
ds_1^2 = -\left(- \half \Lambda x^2 + 2 G_1 M |x| -C\right) dt^2 + \frac{dx^2}{\left(-\half  \Lambda x^2 + 2 G_1 M|x| -C\right)} \label{g11}
\eeq
where $R$ and $T$ are the Ricci and energy-momentum scalars, $G_1$ is the one-dimensional gravitational constant, $r_1 \equiv |x_H|$ and $C$ is an arbitrary constant of integration \cite{mann2}.  The black hole's entropy and Hawking temperature are respectively
\bea
S_1& =& \frac{2\pi}{\hbar}~\ln\left(\frac{\sqrt{G_1^2 \Mbh^2 - \Lambda}+G_1 \Mbh}{M_0}\right) \label{s11} \\
T_1& =& \frac{\hbar}{2\pi} \sqrt{\Mbh^2 - \frac{C\Lambda}{2}} \label{t11}
\eea
Here, $M_0$ is an arbitrary constant of integration with dimensions of mass.  When the product $C\Lambda$ is reasonably small, the temperature runs linearly with the mass: $T_1 \sim \Mbh$.  

In $(d+1)-$dimensions, the relation between the radiative power of a black hole of mass $\Mbh$ and temperature $T_d$ is described by the generalized Stefan-Boltzmann law  \cite{alnes} 
\beq
{\cal P}_d = \left[\frac{dM}{dt}\right]_d = -\sigma_d A_{d-1} T_d^{d+1}~~,
\label{power}
\eeq
where $A_{d-1} \sim r_H^{d-1}$ is the horizon area and $\sigma_d \sim k_{\rm Boltzmann}^{d+1}$.  The black hole decay time is
\beq
\tau_d = \int_{\Mbh}^0 \frac{dM}{{\cal P}_d}~.
\label{lifetime}
\eeq
It is well-known that $d=3$ black holes have a Hawking temperature $T_3 \sim M_{\rm BH}^{-1}$ and emit radiation as ${\cal P}_3 \sim \Mbh^{-2}$.  

There is a pathological issue in $\done$-D that hinders the calculation of (\ref{power}).  The radiative power is a function of the horizon area, which in this case is ill-defined.  It has recently been shown this problem may be circumvented by re-interpreting the $d-$dimensional area in terms of holographic information bits $A_d = N_{\rm bits} G_{d+1}$, where $N_{\rm bits}$ is an intrinsic bit-count on the horizon \cite{jrmrbm1}.  In the case of a two-dimensional black hole, the horizon consists of antipodes and the bit-count is constant.   The energy radiated from a generic $(1+1)-$D black hole with temperature (\ref{minscalar1}) is thus ${\cal P}_1({\alpha}) \sim M^{2\alpha}$, which in the case considered herein is ${\cal P}_1 \sim \sigma_1\ G_1\ M_1^2$.  As expected, more massive PBHs will radiate away quicker than smaller ones, which can subsequently lead to a model-specific population distribution different from $\dthree$-D models. 

In $\dtwo$-D, the conformal tensor vanishes and the Riemann tensor can be written uniquely in terms of $R_{\mu \nu }$ and $R$.  The BTZ metric solution and temperature are \cite{btz1,btz2,carlipbook}:
\bea
ds^2_2  &=& -\left(G_2 M +\Lambda r^2\right) dt^2 + \frac{dr^2}{G_2 M + \Lambda r^2}  \label{g21}\\
T_{2}  &=&  \sqrt{-G_2\ M\ \Lambda} \label{t21}
\eea
where $\Lambda = -\ell^{-2}$ defines the anti-deSitter scale.  Since the temperature (\ref{t21}) and the horizon $r_H = \sqrt{-\frac{G_2 \Mbh}{\Lambda}}$ are explicitly dependent on the cosmological constant, this introduces the curious side-effect that there are no black holes in $(2+1)$-dimensions unless the spacetime is anti-deSitter.  Furthermore, from the parameters given in (\ref{t21}), the lifetime is infinite unless there is a lower cut-off mass for the black hole stemming from quantum gravity effects.

In both cases considered above, new aspects of PBH physics are introduced by the idea of dimensional evolution.  For the two-dimensional case, PBH population distributions can shift due to the radiative power's quadratic mass dependence, which would lead to a fewer large black holes and a higher number of microscopic ones.    In three dimensions, the vanishing of the temperature would halt the evaporation process during this epoch.  Although extremely speculative, a dimensional evolution scenario provides several distinct consequences that could in principle influence early-universe mechanisms that rely on PBH populations. 

\section{PBH remnants from evaporation}
\label{section3}
The end stage of black hole evaporation is not well understood when the horizon size approaches the scale at which quantum gravity
becomes important.  One possibility that cannot be ruled out is that there are stable remnants \cite{remnant1,giddings,piero1,piero2}.  As a $\dthree$-D PBH radiates away, it will shrink to the point where its horizon size becomes commensurate with the length scale $L_2$ at which the spectral dimension reduces.  If $\Lambda \geq 0$, an evaporating PBH will enter this domain and become a remnant with mass $M_{\rm remnant} = L_2/2G_3$.  If this transition occurs at the terascale, one finds $M_{\rm remnant} =10^{32}~$TeV, or $10^8~$kg.   

According to standard black hole thermodynamics, at present all PBHs of mass $\Mbh \leq 10^{12}~$kg will have evaporated.  If these exclusively account for the mass of dark matter $M_{\rm DM}$ (an overly simplistic but straightforward scenario), the total number can be estimated as $N_{\rm PBH} \sim M_{\rm DM} / 10^8~$kg.   The mass of the visible universe is on the order of $10^{52}-10^{54}~$kg  \cite{jrmrbm}, and with a dark matter content of roughly $20\%$ \cite{wmap} it can be deduced there are $N_{\rm PBH} \sim 10^{45}-10^{47}$ such remnants.  

This situation -- evaporating black holes in a $\dthree$-D universe that eventually reach the $\dtwo$-D threshold -- could be called a ``top-down'' evolution process.  What might have happened to PBHs created in an initially lower-dimensional universe, which survived long enough to make the transition to a higher dimension ({\it i.e.} ``bottom-up'' evolution)?  The exact form of the population distribution would depend on the cosmological model employed.  In a standard Friedmann universe, the mass of a PBH created $t$ seconds following the Big Bang is $M \sim \frac{c^3t}{G_3}$ \cite{carr3}, which assumes the event horizon is on the order of the particle horizon.   

If the number of spacetime dimensions is lower in an earlier epoch, this relationship must be modified.   Such calculations are left for future works.  Assuming continuity of the behavior across dimensional transitions, however, one can make some initial statements about the PBH population distribution just prior to the four-dimensional era.     The temperature $t~$seconds after the Big Bang in a purely $(3+1)$-D relativistic model drops as $T(t) \sim 10^{-6} t^{-1/2}~$TeV, and so $t_{\rm TeV}\sim 10^{-12}~$seconds.    The maximum mass of a PBH created at a terascale $(2+1)\rightarrow(3+1)$-D ``transition'' is thus $\Mpbh \sim 10^{23}~$kg, which would evaporate in the standard fashion and still be present in today's universe.  The age of the universe when the  $\done \rightarrow \dtwo$-D shift occurred (at scales of at least 100~TeV \cite{jrmds1}) would be approximately $t_{\rm 100~TeV} \sim 10^{-16}~$seconds, allowing PBHs of mass $M \sim 10^{19}~$kg to have been created at this stage. 


Reference \cite{carlipbh} provides specific insight into the thermodynamics of black holes from the perspective of spectral dimension reduction in CDT-like scenarios.  Consistent with the above conclusion, it is demonstrated that evaporation ceases once the spectral dimension becomes $(2+1)$.  The remnant is defined for observers outside the horizon in the sense that they cannot probe the internal structure of the black hole, and thus cannot observe any further dimensional reduction behavior that may occur at scales smaller than the horizon.  Observers who are interior to the horizon will be able to detect this dimensional reduction, but universally all observers are limited to resolutions no less than $(1+1)$-D.  A full thermodynamical analysis of this model would help shed light on associated PBH creation, evaporation, and population statistics.

\section{PBH remnants from entropy conservation}
The concept of dimensional transition and its effects are not well understood, and are likely highly model-dependent.  Since the thermodynamic properties of black holes depend on the spacetime dimension in which they live, the transition itself may introduce a new type ``remnant.''  Traditionally, this term refers to a non-thermal end-stage of black hole evaporation.  The spirit of this definition is upheld in the mechanism discussed in Section~\ref{section3}.  In the following section, however, ``remnant'' will refer to an object which is a black hole in $d$-dimensional spacetime, but not in $(d+1)$-dimensions.   Although the exact phenomenology arising from a dimensional transition depends largely on the underlying mechanism, rudimentary assumptions can still be made about the behavior of a PBH as it crosses the $d\rightarrow (d+1)$-D boundary.

Let the entropy of the PBH in $d$-dimensions be $S_d$, and the entropy of the ``evolved'' PBH in $(d+1)$-dimensions $S_{d+1}$.   Assuming such evolution is adiabatic, one may conjecture a non-decreasing entropy for the corresponding PBHs, $S_d \leq S_{d+1}$.  A ``dimensional remnant'' in $(d+1)$-dimensions is an object having the same mass $M$ as the $d$-dimensional black hole, but whose entropy $s_{d+1}(M)$ is not maximized according to the area law.  These quantities thus satisfy the general relation
\beq
S_d(M) \leq s_{d+1}(M) < S_{d+1}(M)~~~,
\label{entropycond}
\eeq
where the object is a black hole if (and only if) its entropy is $S_{d+1}$.  Conversely, neither remnants nor black holes form if $S_d > S_{d+1} > s_{d+1}$.


The above prescription requires some elaboration.   Due to the presence of $\Lambda$ in the defining characteristics of $\dtwo$-D black holes, the only possible scenario in which such objects could consistently exist across dimensional transitions is when each spacetime is anti-deSitter.   If $\Lambda \geq 0$, black holes only exist in $\done$-D and $\dthree$-D, but not $\dtwo$-D.  To maintain
BHs in the three-dimensional epoch, a mechanism must be introduced to map $\Lambda \rightarrow -\Lambda$ (provided $\Lambda \ne 0$).  Although no such process is known, a recent proposal suggests a framework for producing an effectively-positive cosmological constant on semiclassical scales from a wavefunction defined in a space with $\Lambda < 0$ \cite{hawking}.   

The spacetimes that contribute to PBH formation and dimensional remnants will therefore be $\done$- and $\dthree$-D.  A PBH created in the former era will survive into the latter provided $S_4 \geq S_2$.  Remnants are created when $s_4 < S_4$ and $s_4 \geq S_2$.  If the entropy condition is not met ({\it i.e.} if $S_2 > S_4$), no PBHs can form.  The remnant must satisfy $S_4 > s_4$, but since this implies $S_2 > s_4$, the process is unphysical (entropy has decreased) and no remnant forms.  

%

\begin{figure}[h]
\begin{center}
\includegraphics[scale=0.6]{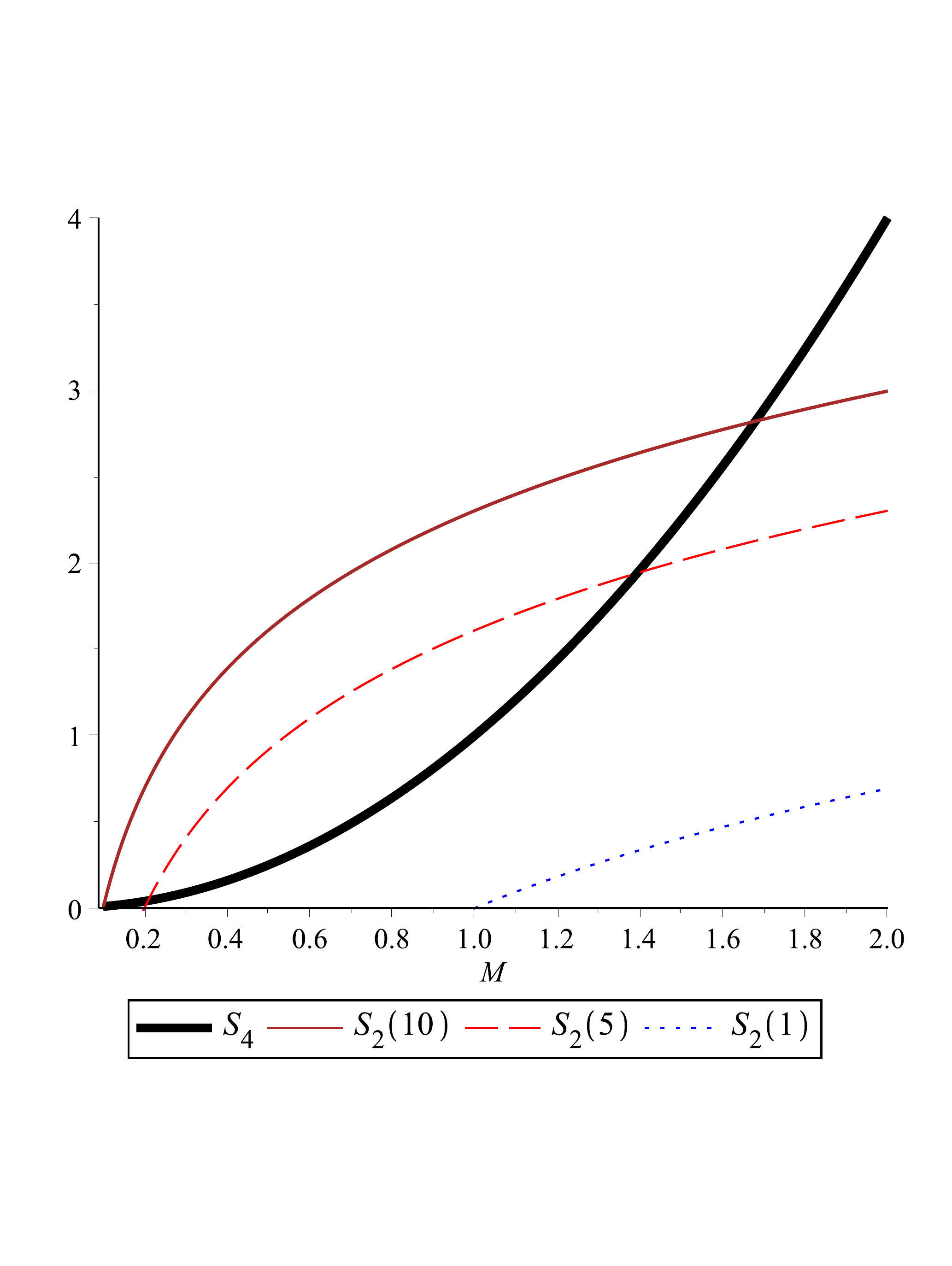}
\caption{
An idealized representation of the gravitational entropies $S_2(\gamma)=\ln(\gamma M)$ and $S_4=M^2$, where $G_3 = \ell_P = \hbar= 1$ and $\gamma=4\pi M_0^{-1}$.  As $M_0$ decreases, a parameter-dependent region is introduced in which the inequality (\ref{s1s3}) is violated.  This threshold occurs at $\gamma \approx 2.332$ ($M_0 \approx 5.39$).}
\label{fig1}
\end{center}
\end{figure}

From the expressions in Section~2, one can explicitly calculate the bound
\beq
S_2 \leq S_4 ~~~\Longrightarrow~~~ \ln\left(\frac{2 G_1 \Mpbh}{M_0}\right) \leq \frac{G_3^2 \Mpbh^2}{\ell_P^2}~~,
\label{s1s3}
\eeq
up to overall constant factors.  Adopting the Myers-Perry definition of the $d-$dimensional gravitational constant,
\beq
G_d=2 \ \pi^{1-\frac{d}{2}}\Gamma\left(\frac{d}{2}\right)\left(\frac{1}{\Mpl}\right)^{d-1}
\eeq
one can assign $G_1 = 2\pi$ and $G_3 = \Mpl^{-2}$.  The above inequality is then
\beq
\ln\left(\frac{4\pi \Mpbh}{M_0}\right) \leq \frac{\Mpbh^2}{\ell_P^2 \Mpl^4}
\eeq
Two situations arise, depending on the value of $M_0$.  First, the  inequality (\ref{s1s3}) is always satisfied, and the PBH mass must be $\Mpbh > \frac{M_0}{4\pi}$ so that $S_2 > 0$.  Second, there may be a range of black hole masses $\Mpbh \in [M_1,M_2]$ (between the intersection points where $S_2=S_4$) for which $S_2 > S_4$  and the general condition (\ref{entropycond}) is violated (see Figure~\ref{fig1}).

The fate of these remnants, as well as any mass distribution that does not meet the proposed criteria (\ref{entropycond}) is unknown and requires further investigation.  The process described in the following section, however, may be one possible result.

\section{PBH electroweak bursts at dimensional transition}
In addition to remnants, dimensional transition may lead to alternate end-stages for PBHs.  When these objects cross into the $\dthree$-D universe, their temperature increases dramatically.  If this exceeds the electroweak symmetry breaking scale of $T_{\rm EW} \sim 200~$GeV, baryon number violating $SU(2)\times U(1)$ processes become unsuppressed and the possibility of electroweak burning exists.  This scenario is similar to the recent proposal in the literature \cite{dsew} of ``electroweak stars,'' in which the EW ``thermal'' pressure balances the inward gravitational collapse of a stellar body.    In this event, the PBH (or its remnant described in Section~4) would evaporate instead in an electroweak burst, whereby quarks are converted into leptons.  Detection of such explosions could therefore provide support for the mechanisms proposed in this letter.

\section{Open Questions and Future Directions}
If Planck-scale physics is indeed set against an effective lower-dimensional background, the consequences are numerous and potentially testable.  The resulting shift in PBH population density may well have an impact on structure formation, if PBHs are dark matter candidates.  A logical future extension of this proposal would address the impact on Reissner-Nordstr\"om and Kerr-Newmann PBHs.    Alternate but critical consideration must be paid to the population statistics and mechanisms of PBH formation in a lower-dimensional arena, including quantum fluctuation characteristics and BH pair production rates \cite{jrmpn3}.

Other outstanding questions remain.  If the proposal \cite{vd1,vd2} is correct and dimensions are indeed ``evolving,'' it is possible the Universe will eventually become $(4+1)$-D.   Has it potentially done so already, and is there evidence to support this contention?  Indeed, such a spacetime at distances on the order of the Hubble length has been suggested \cite{dgp1}, which could act as a potential geometric solution to the dark energy problem \cite{vd1,ref2}.   Since the characteristic length scale exceeds any potential horizon radius, it is perhaps unlikely that this has interesting consequences insofar as black holes are concerned.

Alternatively, evidence of a higher-dimensional spacetime could be imprinted in the large-scale distribution of galaxies.  At least locally, the number density of galaxies $N \sim r^{D_F}$ is well-described as a fractal with $D_F =2$, which is consequently a signature of the distribution's geometry: in this case, it scales as an area.  It has been suggested that this is a holographic-like manifestation of an underlying gravitational theory: the number density of galaxies scales as the boundary of the volume in which they reside, $N(r) \sim \partial V(r)$ \cite{jrmfh}.  The $D_F =2$ fractal scaling does not convincingly extend to the largest of redshifts, however, with transitions to homogeneity ($D_F = 3$) beginning somewhere between $100-1000~$Mpc.  Combining the idea of dimensional evolution with fractal holography, this change in clustering behavior might simply reflect a transition to a higher-dimensional volume.

Lastly, an intriguing consequence of dimensional evolution is the potential observation of fractional dimensions governing gravitational physics.  The notion of a fractal spectral dimension is not new, and some related phenomenology has been considered in the literature.  These include fractional black hole horizon areas \cite{euro2} and ``un''-spectral dimensional reduction \cite{pnes} from a quantum gravitational perspective, as well as the range of quantum field theory modifications discussed in references \cite{calcagni4,cdt,horava,lmpn,calcagni1,calcagni2,calcagni3,calcagni5}.  Detection of non-integer spectral dimensions would certainly lend support to reduction/evolution theories such as those discussed herein.  Probing higher energies may one day reveal such results, provided the transition occurs in a time $t > L_{n,n+1}$ constrained by the energy scale $E_{n,n+1} \sim L_{n,n+1}^{-1}$.

Regardless of the possible dimensional reduction mechanism, the proposals addressed in this letter can ultimately lead to a new and fascinating understanding of primordial cosmology.  If borne out by observation, such key evidence of a dynamical spacetime dimension would represent a tantalizing new perspective on the evolution and fundamental structure of the Universe in which we live.

\vskip 1cm
\noindent {\bf Acknowledgements}\\
\noindent This work was completed in part at the University of Trieste, the Frankfurt Institute for Advanced Studies, and the Perimeter Institute for Theoretical Physics. The author thanks Euro Spallucci, Piero Nicolini, and Niayesh Afshordi for insightful discussions and their hospitality at those respective institutions.  Thanks are also extended to Steve Carlip, Greg Landsberg, Robert Mann, and Dejan Stojkovic for providing helpful suggestions on earlier versions of the manuscript.

%
 \end{document}